\documentclass[manuscript]{aastex}
\begin{document}

\shorttitle{Expanding Shell of Water Masers in W75N}
\shortauthors{Kim et al.}

\title{Evolution of the Water Maser Expanding Shell in W75N VLA~2}

\author{\sc Jeong-Sook Kim$^{1,2}$,
            Soon-Wook Kim$^{2,*}$,
            Tomoharu Kurayama$^{3,4}$,
            Mareki Honma$^5$,
            Tesuo Sasao$^6$,
            Gabriele Surcis$^7$,
            Jorge Cant\'o$^8$,
            Jos\'e M. Torrelles$^9$
            and Sang Joon Kim$^1$,}
\affil{$^1$ School of Space Science, Kyunghee University,
            Seocheon-dong, Giheung-si, Gyeonggi-do, 446-701, Republic of Korea;
            evony@kasi.re.kr.\\
       $^2$ Korea Astronomy and Space Science Institute,
            776 Daedeokdaero, Yuseong, Daejeon 305-348, Republic of Korea;
            skim@kasi.re.kr.\\
       $^3$ Graduate School of Science and Engineering, Kagoshima University,
            1-21-35 Korimoto, Kagoshima, Kagoshima 890-0065, Japan \\
       $^4$ Center for Fundamental Education, Teikyo University of Science,
            2525 Yatsusawa, Uenohara, Yamanashi 409-0193, Japan \\
       $^5$ National Astronomical Observatory of Japan, 2-21-1 Osawa,
            Mitaka, Tokyo 181-8588, Japan \\
       $^6$ Yaeyama Star Club, Ookawa, Ishigaki, Okinawa 904-0022, Japan\\
       $^7$ Joint Institute for VLBI in Europe, Postbus 2, 7990 AA Dwingeloo
            The Netherlands \\
            $^8$ Instituto de Astronom\'{\i}a (UNAM), Apartado 70-264, 04510 M\'exico D. F., M\'exico\\
       $^9$ Instituto de Ciencias del Espacio (CSIC)-UB/IEEC,
            Universitat de Barcelona,
            Mart\'{\i} i Franqu\`{e}s 1, 08028 Barcelona, Spain\\}
\altaffiltext{*}{corresponding author}

\begin{abstract}

We present Very Long Baseline Interferometry (VLBI) observations of 22 GHz H$_2$O masers in the high-mass star-forming region of \objectname{W75N}, carried out with VLBI Exploration of Radio Astrometry (VERA) for three-epochs in 2007 with an angular resolution of $\sim$ 1 mas.
We detected H$_2$O maser emission toward the radio jet in VLA~1 and the expanding shell-like structure in VLA~2.
The spatial distribution of the H$_2$O masers detected with VERA and measured proper motions around VLA~1 and VLA~2 are similar to those found with previous VLBI observations in epochs 1999 and 2005, with the masers in VLA~1 mainly distributed along a linear structure parallel to the radio jet and, on the other hand, forming a shell-like structure around VLA~2.
We have made elliptical fits to the VLA~2 H$_2$O maser shell-like structure observed in the different epochs (1999, 2005, and 2007),
 and found that the shell is still expanding eight years after its discovery.
From the difference in the size of the semi-major axes of the fitted ellipses  in the epochs 1999 ($\simeq$ 71$\pm$1 mas), 2005 ($\simeq$ 97$\pm$3 mas), and 2007 ($\simeq$ 111$\pm$1 mas), we estimate an average expanding velocity of $\sim$
5~mas~yr$^{-1}$, similar to the proper motions measured
in the individual H$_2$O maser features.
A kinematic age of $\sim$ 20~yr is derived for this structure.
In addition, our VERA observations indicate an increase in the ellipticity of the expanding shell around VLA~2  from epochs 1999 to 2007.
In fact, the elliptical fit of the VERA data shows a ratio between the minor and major axes of $\sim$ 0.6, in contrast with a most circular shape for the shell detected in 1999 and 2005 (b/a $\sim$ 0.9).
This suggests that we are probably observing the formation of a jet-driven H$_2$O maser structure in VLA2, evolving from a non-collimated
pulsed-outflow event during the first stages of evolution of a massive
young stellar object (YSO).
This may support predictions made earlier by other authors on this
issue, consistent
with recent magnetohydrodynamical simulations.
We discuss possible implications of our results in the study of the first stages  of evolution of massive YSOs.

\end{abstract}


\section{INTRODUCTION}

   In high-mass star forming regions, H$_2$O masers are locally generated and pumped by
the interaction of supersonic outflows from a central young stellar object (YSO) with a clumpy medium
\citep[e.g.,][]{E92}.
   Since H$_2$O masers are bright and compact sources,
they are one of the major targets for Very Long Baseline Interferometry (VLBI)
with milliarcsecond (mas) angular resolution,
being a powerful tool to know the structure and kinematics of the gas close to massive YSOs.
   In fact,  VLBI and long-term monitoring observations of H$_2$O masers
have provided basic physical information for star-forming regions producing maser emission,
revealing, for example, the internal structure of clusters of masers,
the variability of their flux densities, radial velocities, proper motions,
structure of the magnetic field, and evolutionary stages of massive protostars
\citep[e.g.,][]{Goddi06,S11,B12,Chibueze12,T12,V12}.

   W75N is a massive star-forming region located in the Cygnus X complex of
dense molecular clouds \citep{Dickel78,Shepherd03,Persi06}.
   The detection of H$_2$O, OH, and CH$_3$OH masers in the vicinity of
the radio continuum sources VLA~1, VLA~2, and VLA~3 indicates that
they are massive YSOs in their early stages of evolution, probably early B spectral types
\citep{Baart86,Hunter94,T97,Shepherd04,S09,Carrasco10}.
   These three sources are located within an area of $\sim$ 1.5$''$$\times$1.5$''$
($\sim$ 3000 AU$\times$[d/2~kpc])\footnote{Most of the works published on W75N have been reported
assuming a distance of 2~kpc \citep{Fis1985}.
   However, a new estimate of 1.3 kpc for the distance has been recently
reported through trigonometric parallax measurement \citep{Rygl12}.
   All parameters given in this paper that depend on the distance are
normalized to the distance of 2~kpc for a proper comparison with
results from previous works, but allowing accurate parameter
estimates for the new reported distance of 1.3 kpc.},
with VLA~1 and VLA~3 showing elongated radio continuum emission
interpreted as arising from radio jets, and VLA~2 showing
compact unresolved continuum emission ($\leq$ 0.08$''$) at cm wavelengths
with a still unknown nature (Torrelles et al. 1997,
Carrasco-Gonz\'alez et al. 2010).

   VLBI multi-epoch observations of H$_2$O masers carried out with the
Very Long Baseline Array (VLBA) in 1999 show remarkable different
geometry of outflow ejections in VLA~1 and VLA~2, both objects separated by
$\sim$ 0.7$''$ (1400 AU$\times$[d/2~kpc]; Torrelles et al. 2003, hereafter T03).
   In fact, the H$_2$O masers associated with VLA~1 are distributed
along a linear structure parallel to its radio jet, with mean proper motions of
$\sim$ 2~mas yr$^{-1}$ (19~km~s$^{-1}$$\times$[d/2~kpc]), tracing a collimated
jet-like outflow at scales of $\sim$ 1$''$ (2000 AU$\times$[d/2~kpc]).
   On the other hand, the masers associated with VLA~2 display an almost
circular shell-like structure of $\sim$ 0.08$''$ radius (160 AU$\times$[d/2~kpc])
around the central compact radio continuum source,
moving outward in multiple directions with a mean proper motion of
$\sim$ 3~mas yr$^{-1}$  (28~km~s$^{-1}$$\times$[d/2~kpc])
and estimated kinematic age of 13 yr.
T03 interpret this structure as a wind-driven shell from the central massive YSO.
This result was surprising since the current paradigm of star formation
through accretion disks, ejecting gas via magnetohydrodynamical (MHD) mechanisms,
does not predict outflows expanding without any preferential direction,
but producing collimated outflows \citep[e.g.,][]{Garay99,McKee02}.
On the basis of these VLBA results,
T03 suggested that at the first stages of evolution of massive protostars
there may exist short-lived events with very poorly collimated outflows.
Very recently, this has been theoretically proved
by Seifried et al. (2012), who performed MHD simulations of massive star
formation by including both strong and weak magnetic fields. They found that
at the earliest stage, in the case of a strong magnetic field, the outflows are
poorly collimated for a short-lived period.
   In addition, since the motions of the shell-like structure in VLA~2
are observed at a smaller scale than the collimated motions
in the nearby VLA~1 source,
T03 suggested that VLA~2 is in an earlier stage than VLA~1 and could evolve in the future
into a collimated jet-like structure.
More recently, Surcis et al. (2011; hereafter S11) compared the distribution of the masers around VLA~2 observed in 1999 (T03)
and in a single-epoch of 2005 (S11) and found an average expansion velocity
for the shell-like structure of
$\sim$ 4.8~mas~yr$^{-1}$ ($46~{\rm km}~{\rm s}^{-1}$$\times$[d/2~kpc])
in the time span of six years (1999$-$2005).
   In addition, these authors noted that
there are preferential gas movements northeastward in the VLA 2 shell, suggesting the formation of a jet.

In this paper, we present three epochs of VLBI observations of H$_2$O masers in W75N carried out with the VLBI Exploration of Radio Astrometry (VERA) in 2007.
The main goal of this work is to know the evolution of the expanding shell-like structure of VLA~2 by extending the timescale of the observations (1999-2007).
   In \S 2 we describe the VERA observations and data analysis,
presenting the results in \S 3. In \S 4 we discuss the comparison of our results with those obtained from previous epochs of VLBI observations, focusing primarily on the evolution of the expanding H$_2$O maser shell-like structure of VLA~2.
Their implications in the study of young massive protostars are also discussed. Our main conclusions are presented in \S 5.

\section{OBSERVATIONS AND DATA ANALYSIS}

   The three-epoch of  H$_{2}$O maser line ($6_{16}-5_{23}$, 22235.080 MHz)
observations toward  \objectname{W75N} were carried out on
2007 January 20, 2007 February 21, and 2007 May 29 with VERA,
a Japanese VLBI facility, including all four 20-m antennas.
   The baseline length of VERA ranges from 1,019 km to 2,270 km.
   All VERA antennas have a facility of dual-beam receiving system
\citep{Haschick81,Kawaguchi00}.
   The unique dual-beam mode enables us to carry out more effective
phase-referencing VLBI observation than the conventional antenna
in nodding mode, by simultaneously observing
target and reference sources separated within $2.2^{\circ}$.
   We utilized the dual-beam mode for \objectname{W75N}
by taking the galactic radio continuum source Cygnus X-3.
   However, since the microquasar Cygnus X-3 is a transient source, we only detected
fringes for the first and third epoch during its flaring states.
As a result, it has not been possible to carry out
the phase referencing technique for W75N.

   In each of the three epochs, the total observation time was 8 hours.
   All VERA stations had no systematic and technical problems.
   Left-handed circular polarization was received and
sampled with 2-bit quantization,
and filtered using the VERA digital filter unit \citep{Iguchi05}.
   The data were recorded onto magnetic tapes at a rate of 1024 Mbps,
providing a total bandwidth of 256 MHz.
   We chose the VERA 7 mode in which one intermediate frequency (IF) was assigned to \objectname{W75N}
and the other 15 IFs were assigned to Cygnus X-3.
   The well-known, bright continuum source
BL Lac was observed every 70 minutes for the bandpass and delay calibrations.
   The system noise temperatures including atmospheric attenuation,
sensitive to weather conditions and the elevation angle of the observed source,
were measured with the chopper-wheel method \citep{Ulich76}.
   The aperture efficiencies of the antennas ranged from 45\% to 52\%, depending on each VERA station
(see the VERA status report
\footnote{http://veraserver.mtk.nao.ac.jp/restricted/CFP2010/status10.pdf}).
   A variation of the aperture efficiency of each antenna
as a function of the elevation angle was confirmed to be less than 10\%
at the lowest elevation in the observations around $\sim$ 20$^{\circ}$.

   Correlation processing was performed on the Mitaka FX Correlator
\citep{Chikada91}
located at the National Astronomical Observatory of Japan (NAOJ) Mitaka campus.
   The accumulation period of the correlation was set to be 1 second.
   In the process of correlation, the bandwidth of the IF for W75N
was reduced from 16 MHz to 8 MHz for H$_{2}$O maser line, divided by 512 channels
of 15.625 ${\rm kHz}$ width, corresponding to a velocity resolution of 0.21 ${\rm km~s}^{-1}$.
Therefore, the total velocity coverage is 107.5 ${\rm km~s}^{-1}$,
centered at the local standard of rest velocity V$_{LSR}$ = 12.3~km~s$^{-1}$.

   Calibration and imaging were performed using the Astronomical Imaging Processing System ($\rm AIPS$) of
the National Radio Astronomy Observatory (NRAO).
   The amplitude calibrations were performed based on the system noise temperatures during the observations.
   We used this data for the bandpass calibrations and
the determination of clock and clock-rate offset.
   The Doppler shift due to the Earth's motion
was corrected with the AIPS spectrum channel correction task, CVEL.
   The observed frequencies of the maser lines
in each channel were converted to
radial velocities with respect to the local standard of rest
(LSR) using the rest frequency of 22.235080 GHz for the
H$_{2}$O $6_{16}-5_{23}$ transition.
   For the three epochs of observations, the cross-correlation spectra of
H$_{2}$O maser are shown in Figure 1.
   The cross-power spectra of all VERA baselines were checked
to choose a single frequency channel of emission with unresolved structure
and without rapid amplitude-time variation.
   The fringe fitting process was carried out using the task FRING
with an option of no delay search for the selected frequency channel.
   With the task IMAGR, a synthesized image was made from fringe-fitted data
with a 0.1 mas pixel for a region of 1024$\times$1024 pixels,
or $\sim$100$\times$100 ${\rm mas}^{2}$.
   Phase and amplitude self-calibrations were carried out
based on a synthesized image as an initial model:
a strong, point-like H$_2$O maser
identified at $\rm V_{LSR}$ $\simeq$ 7.9 ${\rm km~s}^{-1}$
and associated with W75N~VLA~2.
   For this maser we estimated absolute coordinates:
$\alpha(J2000)$ = {\rm 20$^h$38$^m$36.49$^s$} ($\pm$0.01$^s$),
$\delta(J2000)$ = 42$^\circ$37$\arcmin$34.3$\arcsec$ ($\pm$0.1$''$).

   As a result of the phase and amplitude self-calibrations,
we obtained the complex gains that we applied to the rest of frequency channels
to produce the phase and amplitude-calibrated synthesized images
in each channel. The resulting synthesized beams were 1.2$\times$0.8 (first epoch),
1.2$\times$0.9 (second epoch), and
1.4$\times$0.9 (third epoch) in units of mas.
   To search for H$_2$O
maser emission, a region of ${2\arcsec}{\times}{2\arcsec}$
was examined including VLA~1, VLA~2, and VLA~3.
H$_2$O maser emission was detected toward VLA~1 and VLA~2,
with typical rms values of $\sim$ 2, 0.9, and 0.3 Jy beam$^{-1}$
for the first, second, and third epoch, respectively.
   In the first epoch, the relatively higher rms value is due to
high system temperatures of $\sim$1,000 K and $\sim$800 K
in the Iriki and Ishigaki stations, respectively.
    To identify the H$_2$O masers, we adopted a signal-to-noise ratio $\ge$ 6
as the detection threshold.
The maser emission occurring at a given velocity channel and position
(hereafter referred as maser spot) was fitted by two-dimensional elliptical
Gaussian using the task JMFIT of AIPS to obtain necessary information
such as the position, flux density, and radial velocity.
   The accuracy in the relative positions of the maser spots at each epoch
is better than $\sim$ 0.1~mas
(estimated from the beam size and signal-to-noise ratio of the emission).

   To measure the proper motions of the H$_2$O masers,
an initial, preliminary
alignment of the three epochs was made with respect to
the reference maser spot used for self-calibrating the data.
   Within this reference system, we measured the proper motions of
three masers that persisted in all three epochs
through a linear fitting of their positions,
and of 37 masers that appeared in two different epochs.
    To consider  the persisting masers
we applied the criterion of having the same radial velocity
($V_{LSR}$) in the different epochs and moving
with proper motions
within $\sim$ 10 ${\rm mas~yr}^{-1}$ (100~km~s$^{-1}$$\times$[d/2~kpc]).
   However, the selected reference maser spot may have its own proper motion,
introducing an arbitrary offset to the proper motions of all the masers.
   To minimize this effect,
we subtracted the mean proper motion vector of all these 40 masers
from the proper motion of each maser,
making a realignment of the three epochs.
   The proper motion values of the 40 masers that we present in Table 1
have been obtained after this realignment.

\section{RESULTS}

   We have detected two clusters of H$_2$O masers,
one associated with VLA~1 and the other with VLA~2.
   No maser emission was detected toward VLA~3.
   In Figure 2, we present the spatial distribution of the masers
in VLA~1 and VLA~2 as observed with VERA for three epochs in 2007.
Globally, the maser distribution is similar to that previously
reported with VLBA in 1999 (T03) and 2005 (S11).
   In fact, while the H$_2$O masers in VLA~1 are mainly distributed
along a linear structure parallel to its radio jet,
the masers in VLA~2 are distributed in a shell-like structure
surrounding the compact radio continuum source.
   The total H$_2$O maser spots  detected with VERA
are 31, 106, and 141, with most of the masers belonging to
VLA~2, 25/31(80 $\%$), 95/106 (90 $\%$) and 92/144 (64 $\%$)
on 2007 January 20, 2007 February 21, and 2007 May 29, respectively.
   The probable reason for the relatively small number of detected masers
in the first epoch is the higher rms level in this set of data in comparison
with the other two epochs (see \S~2).
   As a result, the shell-like structure associated with VLA 2
is not clearly seen in the first epoch (Figure 2).

   The H$_2$O maser emission in VLA 1 spans a radial velocity range
from  $\rm V_{LSR}$ $\simeq$ 6 to 13 ${\rm km~s}^{-1}$,
while that in VLA~2 is in the range of 0$-$16 ${\rm km~s}^{-1}$.
   We do not detect H$_{2}$O maser emission with $\rm V_{LSR}$ lower
than 0 ${\rm km~s}^{-1}$,
which was detected in the two previous VLBA observations
in 1999 (T03) and 2005 (S11),
probably due to the high flux density variability of the H$_2$O masers.
   With respect to the proper motion of the H$_2$O masers,
we estimate the mean values of
$\sim$ 2.1 and 2.5 ${\rm mas~yr}^{-1}$
($\sim$ 20~km~s$^{-1}$$\times$[d/2~kpc] and
$\sim$ 24~km~s$^{-1}$$\times$[d/2~kpc])
in VLA~1 and VLA~2, respectively (Figures 3 and 4; Table 1).
   These mean values are very similar to those obtained previously
with the VLBA observations in 1999, $\sim$ 2 and 3 ${\rm mas~yr}^{-1}$,
respectively (T03).
In \S 3.1 we will concentrate on the H$_2$O maser shell-like
structure detected around VLA~2 based on an
elliptical fitting analysis,
comparing our results with those from previous observations (T03, S11).

\subsection{Ellipticity of the Shell-like Structure around VLA~2}

From Figure 4, we see that the northeastern and southwestern parts of the shell-like structure as observed with VERA (epoch 2007) present
proper motions mainly toward the northeastern and southwestern directions, respectively,
with similar proper motion values ($\sim$ 3 mas~yr$^{-1}$).
The maximum angular separation from the northeastern and southwestern parts of the shell
(those indicated in Figure 4 by the labels B and D) is $\sim$ 220 mas,
while the maximum angular separation within the shell observed in 1999 was $\sim$ 154~mas (T03).
This increase in the angular separation between the tips of the shell
in eight years (from 1999 to 2007)
corresponds to an expanding velocity of $\sim$ 4~mas~yr$^{-1}$ from the center.
This expanding velocity is
consistent with the proper motions measured in the individual masers of our three epochs of VERA observations in 2007 (Figure 4; Table 1).

To perform a more detailed study of the evolution of the shell-like structure,
we have made elliptical fits to the H$_2$O maser positions detected with the VLBA
in 1999 (April 2, May 7, and June 4; hereafter epoch 1999.3) and 2005 (November 21; hereafter epoch 2005.9),
and with VERA (January 20, February 21, and May 29; hereafter 2007.2).
The ellipses have been fitted by minimizing the chi-squared distribution of the distance
measured from the position of the masers to the ellipse, along a line connecting the maser
to the center of the ellipse.
The main parameters of the elliptical fits, i.e. the semi-major (a) and
semi-minor axes (b), and the position angle, are listed in Table 2.
   In Figure 5 we show the position of the masers and the resulting fitted ellipses
for epochs 1999.3 (ellipse T99), 2005.9 (ellipse S05), and 2007.2 (ellipse K07),
plotted together assuming the same center position.
We note that the parameters of the elliptical fits we list in Table 2 differ slightly from those listed
by S11 for epochs 1999.3 and 2005.9 due to a fault in the  algorithm
used for the fitting in that previous paper.

Comparing the ellipses,
we see that the semi-major axis has increased progressively
during the time span of eight years,
from 71$\pm$1 mas in 1999.3 (ellipse T99),
to 97$\pm 3$ mas in 2005.9 (ellipse S05),
and to 111$\pm$1 mas in 2007.2 (ellipse K07).
   This corresponds to expanding velocities in the intervals
T99$-$S05: 3.9$\pm 0.8$ mas~yr$^{-1}$,
S05$-$K07:  10.8$\pm 3.1$ mas~yr$^{-1}$, and
T99$-$K07:  5.1$\pm 0.2$ mas~yr$^{-1}$.
   All these values for the expanding velocities are compatible between them
for the given errors we estimated,
although showing some indications that the shell has suffered an
acceleration in the S05$-$K07 interval (see Figure 6).
   New epochs of VLBI H$_2$O maser observations are necessary to clearly
confirm it.

Another relevant result is that
the expanding H$_2$O maser shell-like structure has increased its ellipticity.
   In fact, the ratio of semi-minor to semi-major axes
of the fitted ellipse (b/a) for epoch 2007.2 is $\sim$ 0.6,
which is smaller than the one derived for the almost
circular shells
observed in epochs 1999.3 and 2005.9, b/a $\sim$ 0.9 (see Figure 5).
   In addition, a more dominant axis of expansion appears
along the northeast-southwest direction in
epoch 2007.2 (PA $\simeq$ 45$^{\circ}$; Figures 4 and 5).
   We discuss the implications of our results in the study of
the first stages of evolution of massive protostars in the next section.

\section{Discussion}

In this section, we focus on
the long-term evolution of the H$_2$O maser shell-like structure associated with the massive YSO VLA~2. Our VERA observations
have extended the time scale in which this structure has been observed, by adding a third long-term epoch (2007.2) to the observations
of T03 (epoch 1999.3) and S11 (epoch 2005.9), providing a solid result that the
shell is still expanding in the plane of the sky at a velocity of $\sim$ 5~mas~yr$^{-1}$ (47~km~s$^{-1}$$\times$[d/2~kpc])
eight years after its discovery.

The kinematic age of this expanding structure is $\sim$ 20 years up to the epoch 2007,
a value in agreement with the kinematic age of 13 years estimated from the VLBA observations of epoch 1999.3 (T03), considering
the eight years between these observations and our VERA observations. As discussed by T03, the most probably scenario for
the origin of this H$_2$O maser structure is a wind-driven shell, for which they obtained
a wind mass-loss rate $(\dot M_w/M_\odot~{\rm yr}^{-1}) = 8\times 10^{-7}(n/10^8~{\rm cm}^{-3})^{1/2}(T/500~{\rm K})^{1/2}$, terminal wind velocity
$(V_w/{\rm km~s}^{-1})$ $=$ $100(n/10^8~{\rm cm}^{-3})^{1/2}(T/500~{\rm K})^{1/2}$
(where $n$ and $T$ are the particle density and temperature of the shell, respectively),
and density outside the shell (molecular gas environment)
$(n_0/{\rm cm}^{-3}) = 2.6\times10^5(n/10^8~{\rm cm}^{-3})$.
These are reasonable parameters for early B stars ($\dot M_w$, $V_w$) and molecular core ($n_0$).
   T03 and S11 also proposed that the ejected wind from the central massive object
should be non-collimated (to explain the almost circular shape of the shell observed in epochs 1999.3 and 2005.9),
as well as repetitive (to explain the short kinematic age of the shell).

Besides that, the H$_2$O maser structure continues to expand.
Our VERA observations also indicate an apparent increase with time in the ellipticity
of the shell, from an almost circular shape as observed in epochs 1999.3 and 2005.9 (b/a $\sim$ 0.9),
to a more elliptical shape as observed in epoch 2007.2 (b/a $\sim$ 0.6),
with a more dominant expansion along an axis with PA $\sim$ 45$^{\circ}$ (see \S 3, Figure 5 and Table 2).
This result is consistent with the
prediction made previously by S11, who noted that in the epoch 2005.9 there are preferential gas movements northeastward,
suggesting the formation of a collimated outflow in that direction
(see H$_2$O maser positions of S11 at $\sim$ [120, 80] mas in Figure 5).
Furthermore, we found out that the previous proper motion measurements of the H$_2$O masers of the VLA~2 shell in epoch 1999.3
were already showing larger velocities in the plane of the sky along the northeast-southwest direction (see Figure 1 of T03).
   In VideoT03 (Figure 7; published online), we show the expanding motions of the shell as measured by
T03 but extrapolated 10 years ahead starting in epoch 1999.3, where these main motions along the northeast-southwest direction are clearly seen.
   We think that this gives even more support to the northeast-southwest expansion direction of the shell in VLA~2.

Based on these results, we suggest that we are probably observing the formation of a
collimated jet-driven H$_2$O maser structure, evolving from a non-collimated pulsed-outflow event during the first stages of evolution of a massive YSO. The position angle of the new forming jet in VLA~2 would be $\sim$ 45$^{\circ}$,
which interestingly is
similar to the position angle of the nearby more evolved VLA~1 radio jet (PA $\sim$ 40$^{\circ}$; T03), being separated both YSOs by $\sim$ 0.7$''$
($\sim$ 1400 AU$\times$[d/2~kpc]).
In addition, this position angle is close to the direction of
the large-scale magnetic field aligned to the large-scale bipolar
molecular outflow in the region
($\sim$ 5$'$, or $\sim$ 3 pc, PA $\sim$ 70$^{\circ}$; Shepherd et al. 2003; S11),
but it is far from the average direction of the local magnetic field around VLA~2
($\phi_{\rm{B}}^{\rm{VLA2}}\sim18^{\circ}$; S11).
Actually, the local magnetic field around VLA~2 shows
a radial morphology consistent with the recent simulations of
massive star formation \citep[e.g.,][]{Seifried12}.
This suggests that the local magnetic field in the region plays an
important role in the early phases of the formation of the outflows
but afterwards the large-scale magnetic field starts to dominate also in VLA~2.
Even though, S11 measured a stronger magnetic field around VLA~2
($|B|_{\rm{max}}^{\rm{VLA2}}\approx1000$~mG) than around VLA~1
($|B|_{\rm{max}}^{\rm{VLA1}}\approx800$~mG).
To explain the evolution from a non-collimated outflow event to a collimated outflow in VLA~2,
we speculate on the presence of a disk-like gas structure around this massive object,
oriented southeast-northwest (perpendicular to the orientation of the local magnetic field in the region).
In this scenario, an ``isotropic'' ejection event from the central source would expand more freely
in the direction perpendicular to the disk-like structure, along the magnetic field lines.
If this scenario works, it would imply that the disk-like structure has a central ``hole" (cavity) of $\sim$ 60$-$90 mas radius,
the size of the semi-minor axis of the shell.
This scenario fits quite well with the three-dimensional MHD simulations performed by \citet{Seifried12}.
They found that in the earliest stage of a massive star formation,
the outflows are non-collimated if a strong magnetic field is present.
Afterwards, their collimation increases quickly due to the development
of a fast, central jet coupled to the build-up of a Keplerian disk.

``Short-lived'' pulsed ejections (some of them isotropic or non-collimated) have been observed through VLBI H$_2$O maser observations in different massive protostars, including W75N VLA2.
It is still unclear whether these are non-standard phenomena in particular types of sources, or if all massive protostars undergo short-lived pulsed ejection phases. Nevertheless, their presence provides new insights in the study of
the formation of high-mass stars
\citep[e.g.,][]{T01,T11,Sa12,Tri13}.
What we are observing now in W75N VLA~2 is the likely formation of a collimated outflow from a non-collimated expanding shell,
as theoretically expected from the MHD simulations of \citet{Seifried12}.
However, we think that new epochs of VLBI H$_2$O maser observations,
even in full polarization,
are mandatory to confirm this tendency found with our VERA observations (including a possible acceleration in the shell).
JVLA observations at cm/mm wavelengths would be very important to see if the central radio continuum source (VLA~2)
has also evolved into an elongated
structure along the northeast-southwest direction,
like the nearby radio jet in VLA~1.
High-angular resolution, sensitive
(sub)mm observations to study the dust and molecular gas distribution around VLA~2 would
also help to understand the evolution of this expanding shell.

\section{Conclusions}

   We have carried out three epochs of VLBI observations of H$_2$O masers
during 2007 toward the high-mass star-forming region of W75N with
$\sim$ 1 mas of angular resolution, comparing our data with previously
obtained in 1999 and 2005.
   We find that the H$_2$O maser shell associated with the massive YSO VLA~2
is still expanding at $\sim$ 5~mas~yr$^{-1}$
($\sim$ 47~km~s$^{-1}$$\times$[d/2~kpc])
eight years after its discovery in epoch 1999.
   The shell, with a current size of 222~mas$\times$136~mas (444~AU$\times$272~AU at 2~kpc)
and kinematic age of $\sim$ 20~yr,
shows a ratio between the minor and major axes of $\sim$ 0.6,
in contrast with a most circular shape of the shell as observed in 1999 and 2005.
   These results would suggest that we are probably
observing the formation of a jet-driven water maser structure, as
previously proposed and consistent with recent MHD simulations of
massive star formation by Seifried et al. (2012).
   These MHD simulations predict that at the earliest stage,
the outflows are poorly collimated for a short-lived period,
evolving into a collimated outflow, as we are probably observing in VLA~2.
   New epochs of VLBI H$_2$O maser observations are mandatory to confirm
this tendency found with our VERA observations (including a possible acceleration in the shell).

\bigskip

\acknowledgments
We would like to thank our referee for his/her very careful review
of the manuscript, which has helped to substantially improve its
content.  J.C. acknowledges support from CONACyT (M\'exico) grant
61547.  S.J.K. acknowledges partial supports from the Korea Science
and Engineering Foundation (R01-2008-000-20002-0) and a WCU Grant
(No. R31-10016).  J.M.T acknowledges support from MICINN (Spain)
AYA2011- 30228-C03 (co-funded with FEDER funds) and AGAUR (Catalonia)
2009SGR1172 grants. We acknowledge Miquel Trias for helpful comments
on the manuscript.

\clearpage

\begin{figure}
\includegraphics[scale=0.6]{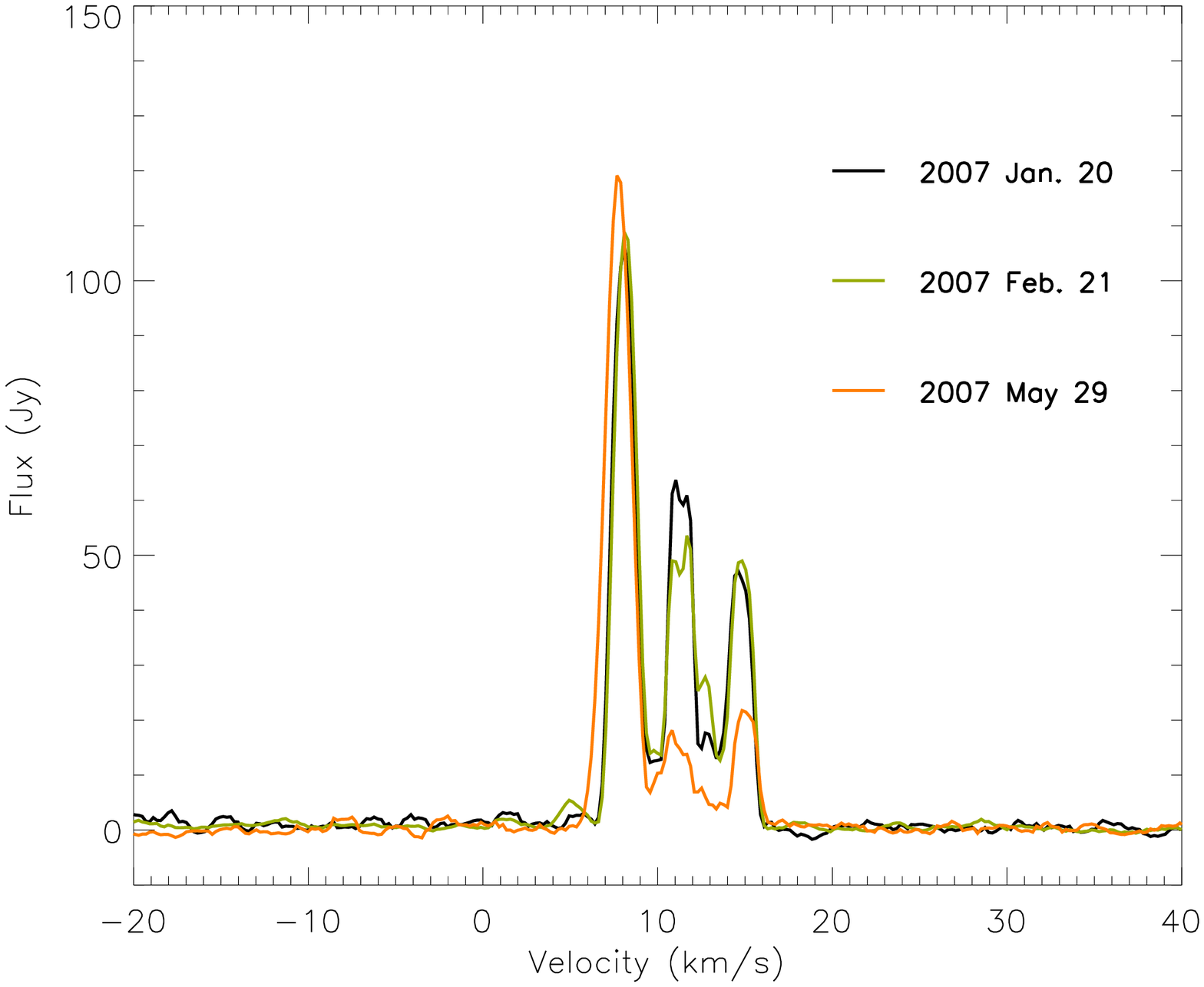}
\bigskip
\caption{Example of scalar-averaged cross power spectra
of W75N observed with the VERA Mizusawa-Iriki baseline (1,276 km).
The black, green, and orange solid lines represent
the observations on 2007 January 20, 2007 February 21, and 2007 May 29,
respectively.
\label{fig1}}
\end{figure}
\clearpage

\begin{figure}
\includegraphics[scale=0.52]{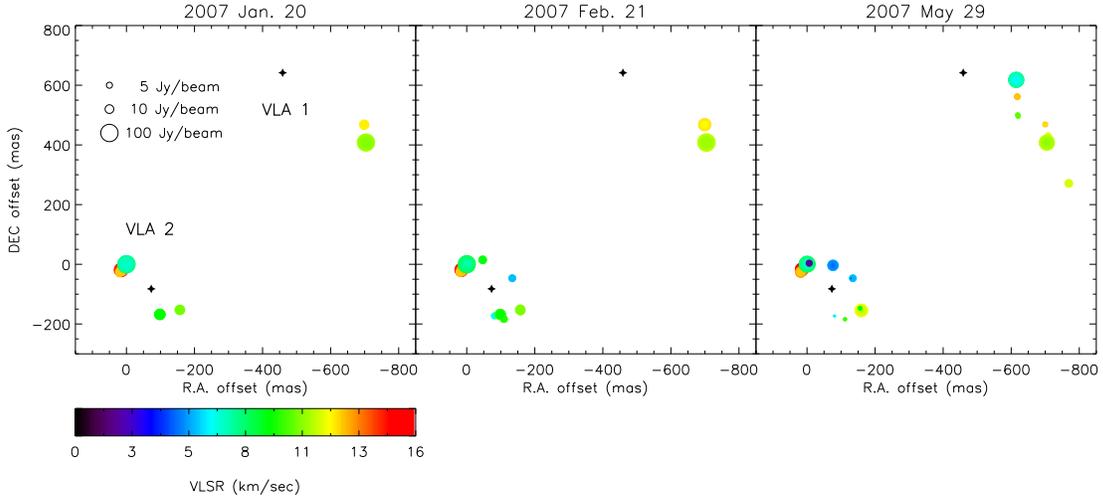}
\bigskip
\caption{Position of the H$_{2}$O maser spots (filled circles)
observed with VERA at 22 GHz
on 2007 January 20 (left), February 21 (middle), and May 29 (right panel)
around the VLA~1 and VLA~2 radio continuum sources (indicated by plus symbols).
The filled circle symbols of the masers are scaled logarithmically
according to their peak flux density.
Open circle symbols for  5, 10, and 100 ${\rm Jy~beam}^{-1}$ are
shown in the upper left corner of the first panel.
The color of the symbols are codified by their radial velocity.
The offset positions are relative to the reference maser spot position (0, 0)
used for self-calibrating the data.
For this maser we estimate absolute coordinates:
$\alpha(J2000)$ = {\rm 20$^h$38$^m$36.49$^s$} ($\pm$0.01$^s$),
$\delta(J2000)$ = 42$^\circ$37$\arcmin$34.3$\arcsec$ ($\pm$0.1$''$).
\label{fig2}}
\end{figure}

\begin{figure}
\epsscale{.80}
\plotone{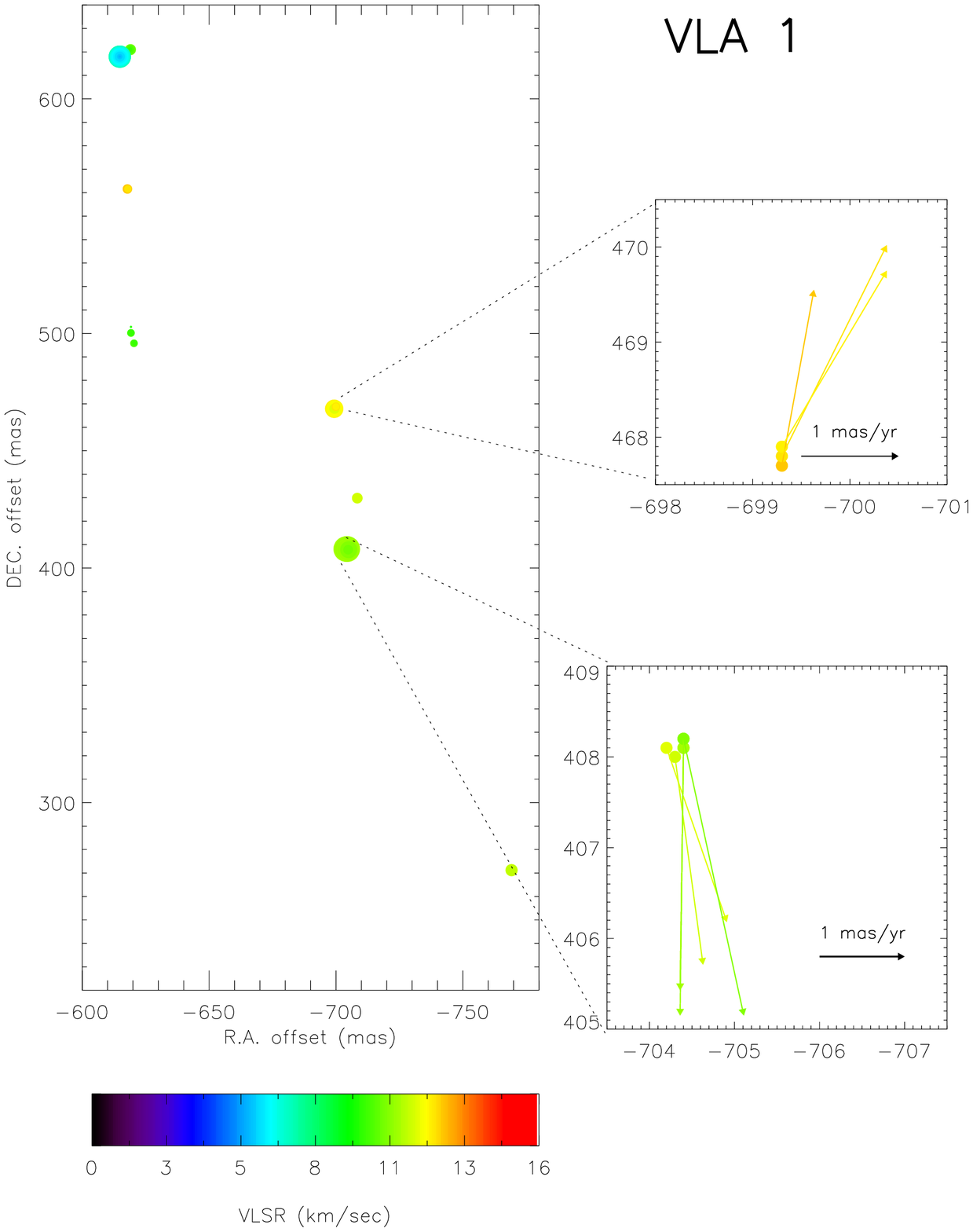}
\caption{
Position of the H$_{2}$O maser spots for all three epochs and
measured proper motions of the masers in VLA 1.
The vectors indicate the direction and values of the proper motions
listed in Table 1.
   Sizes of filled circles and color codes are the same as those
shown in Figure 2.
   Two groups of masers are shown as close-up.
\label{fig3}}
\end{figure}
\bigskip
\clearpage

\begin{figure}
\epsscale{.90}
\plotone{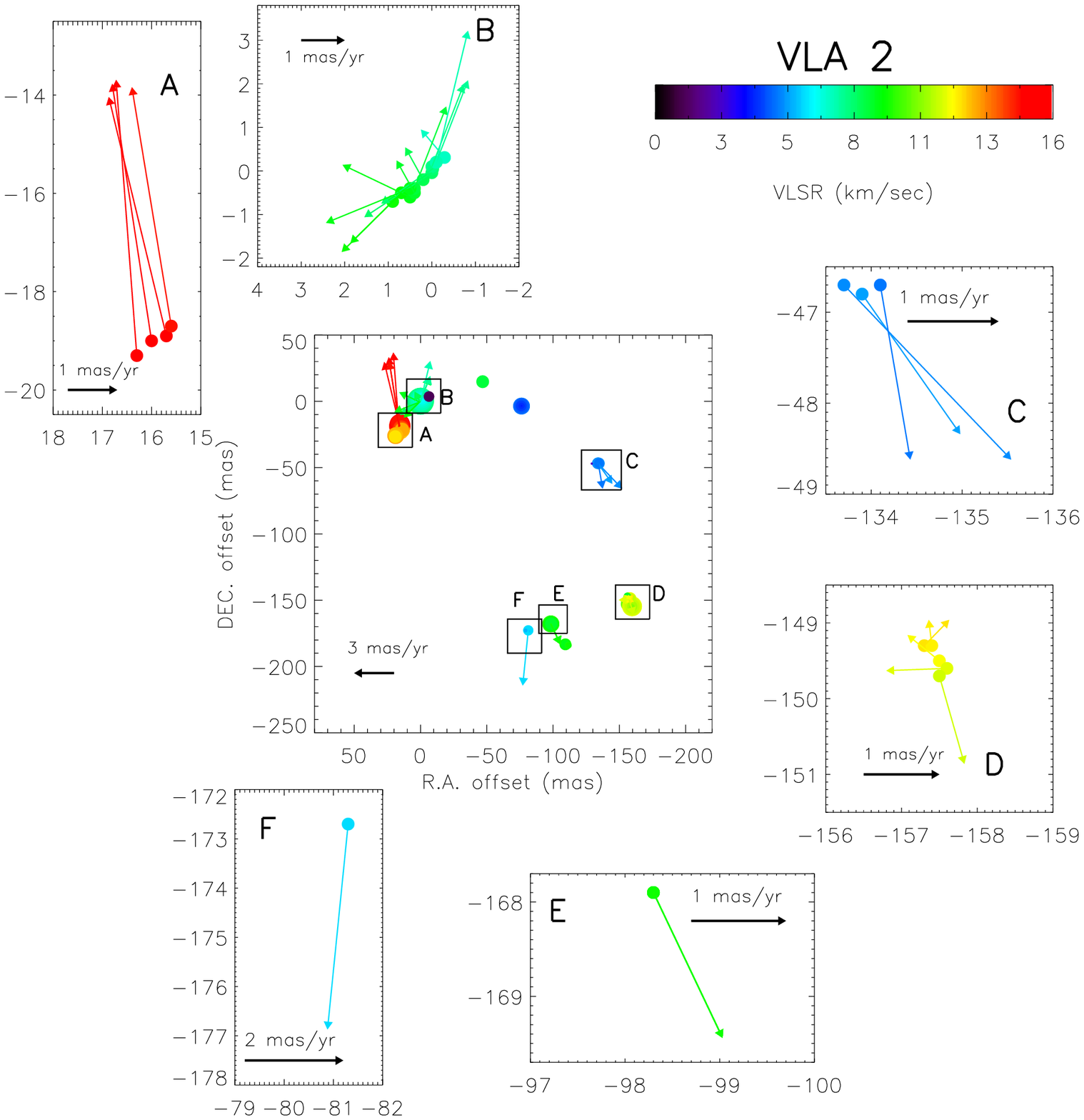}
\caption{
Position of the H$_{2}$O maser spots for all three epochs and
measured proper motions of the masers in VLA 2.
The vectors indicate the direction and values of the proper motions
listed in Table 1.
   Sizes of filled circles and color codes are the same as those show in Figure 2.
Six groups of masers (A, B, C, D, E and F)
are shown as a close-up.
\label{fig4}}
\end{figure}
\bigskip
\clearpage

\begin{figure}
\epsscale{.90}
\plotone{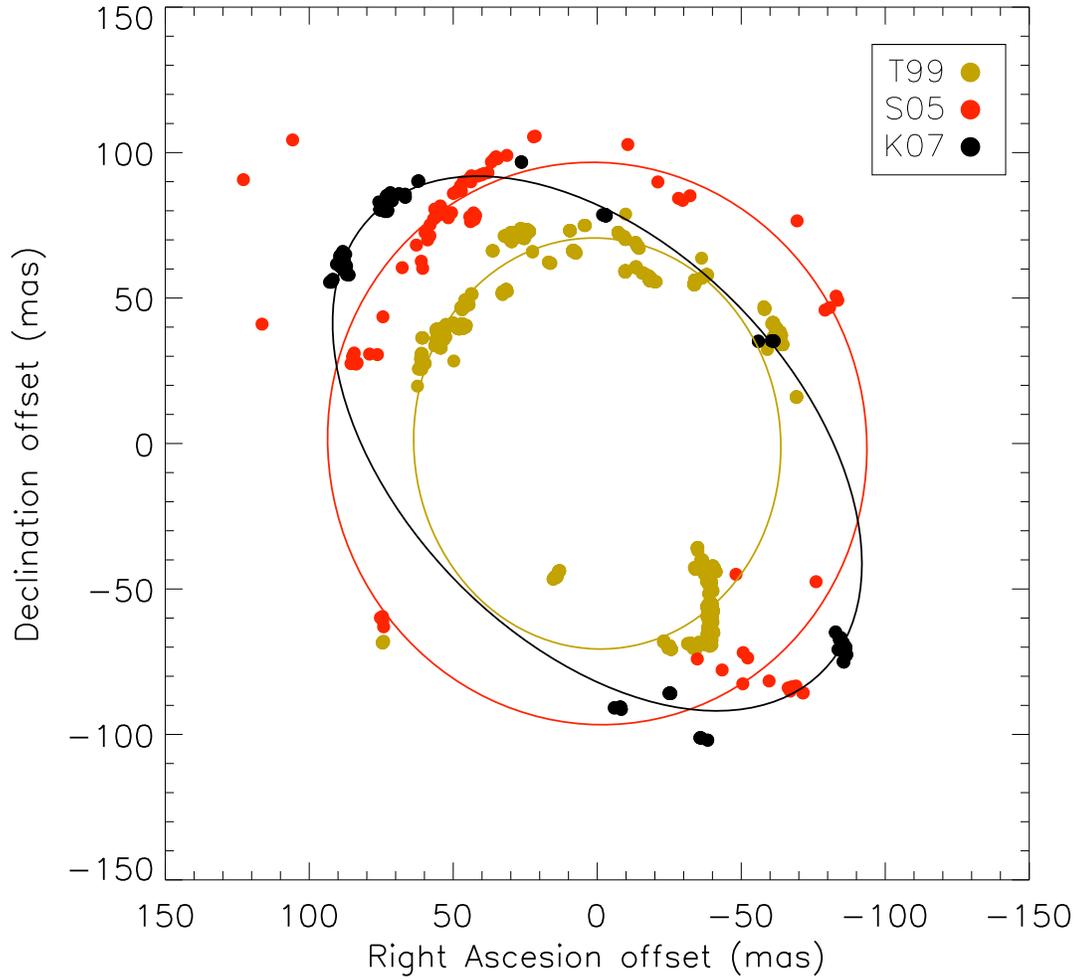}
\caption{Elliptical fits for the H$_2$O masers in VLA~2 observed with VLBI
in epochs 1999.3 (T99), 2005.9 (S05), and 2007.2 (K07) by Torrelles et al. (2003), Surcis et al. (2011), and this work, respectively.
  \label{fig5}}
\end{figure}
\clearpage

\begin{figure}
\includegraphics[angle=90,scale=0.6]{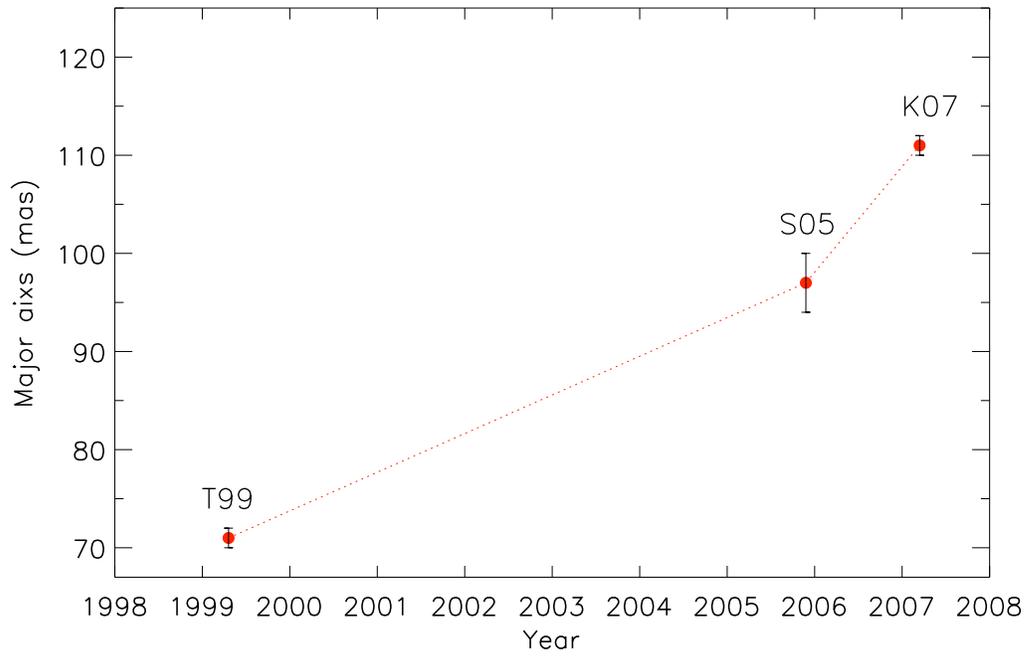}
\caption{Evolution of the size of semi-major axis of the fitted ellipses to the expanding H$_2$O maser shell in VLA 2 for epochs 1999.3, 2005.9, and 2007.2 (Table 2).
\label{fig6}}
\end{figure}

\begin{figure}
\epsscale{.90}
\plotone{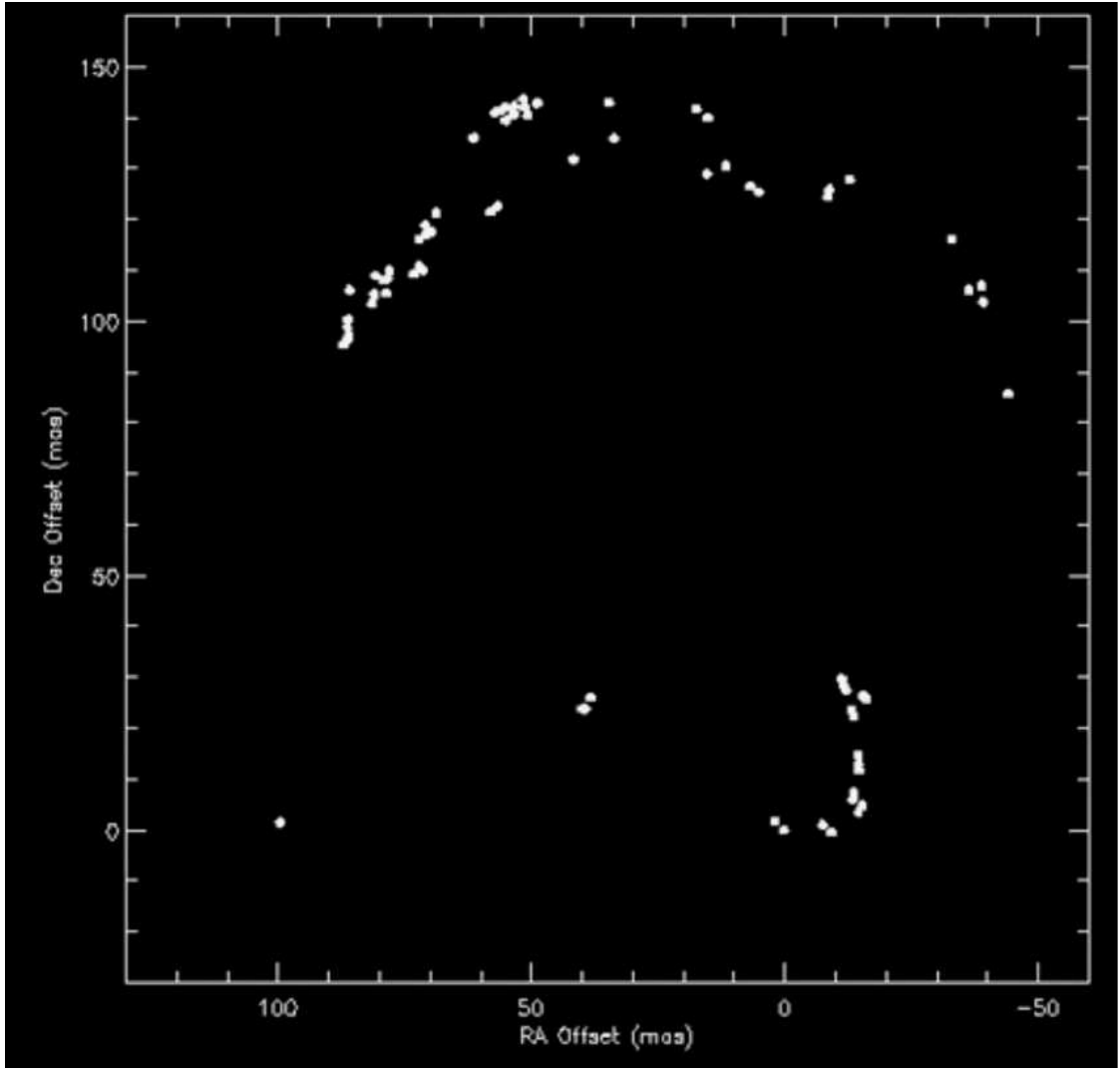}
\caption{Video showing the expanding motions of the shell as measured by T03 but extrapolated 10 years ahead starting in epoch 1999.3. Dots represent the position of the water masers detected by T03.
\label{fig7}}
\end{figure}

\begin{figure}
\includegraphics[scale=0.7]{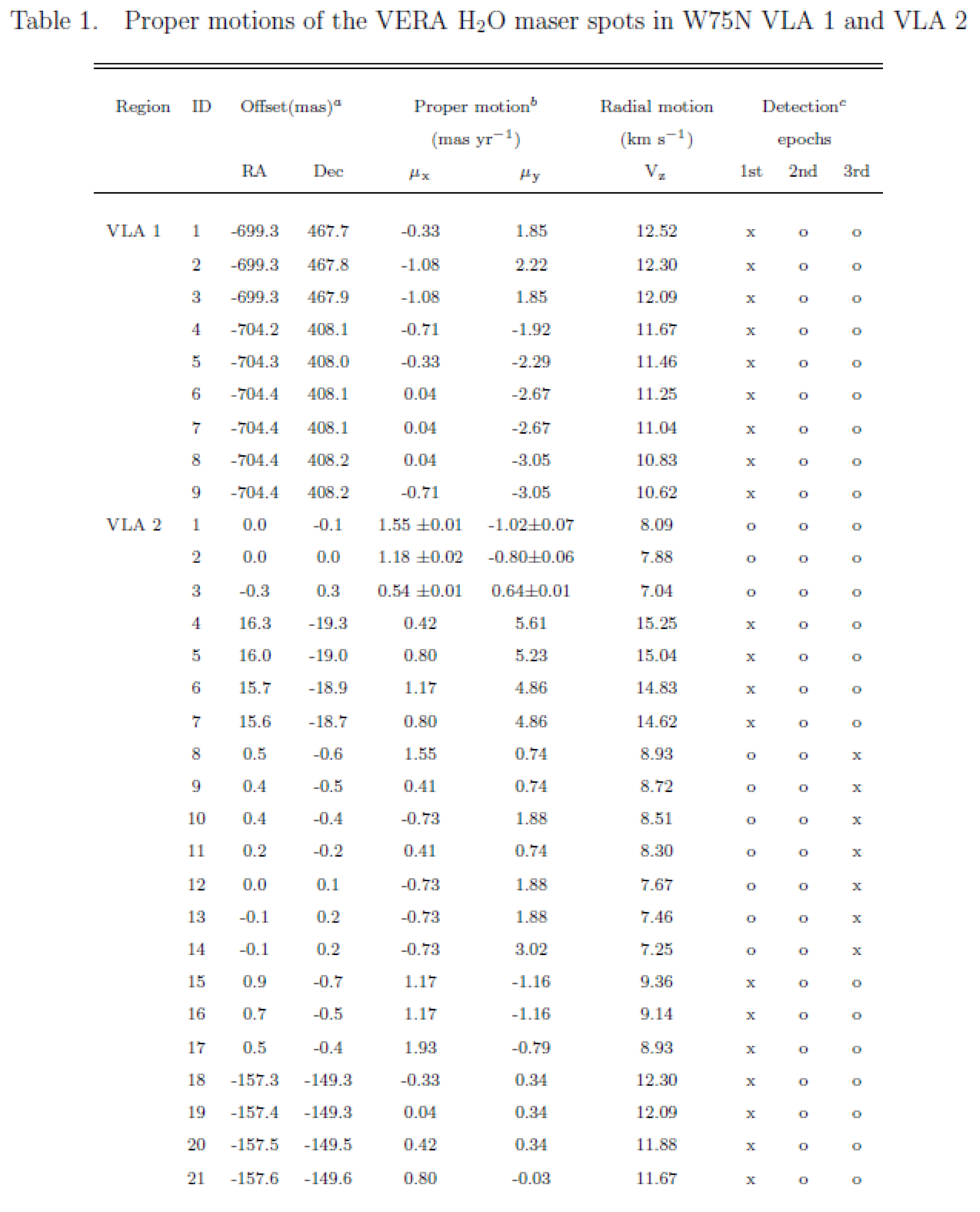}
\bigskip
\end{figure}
\clearpage

\begin{figure}
\includegraphics[scale=0.7]{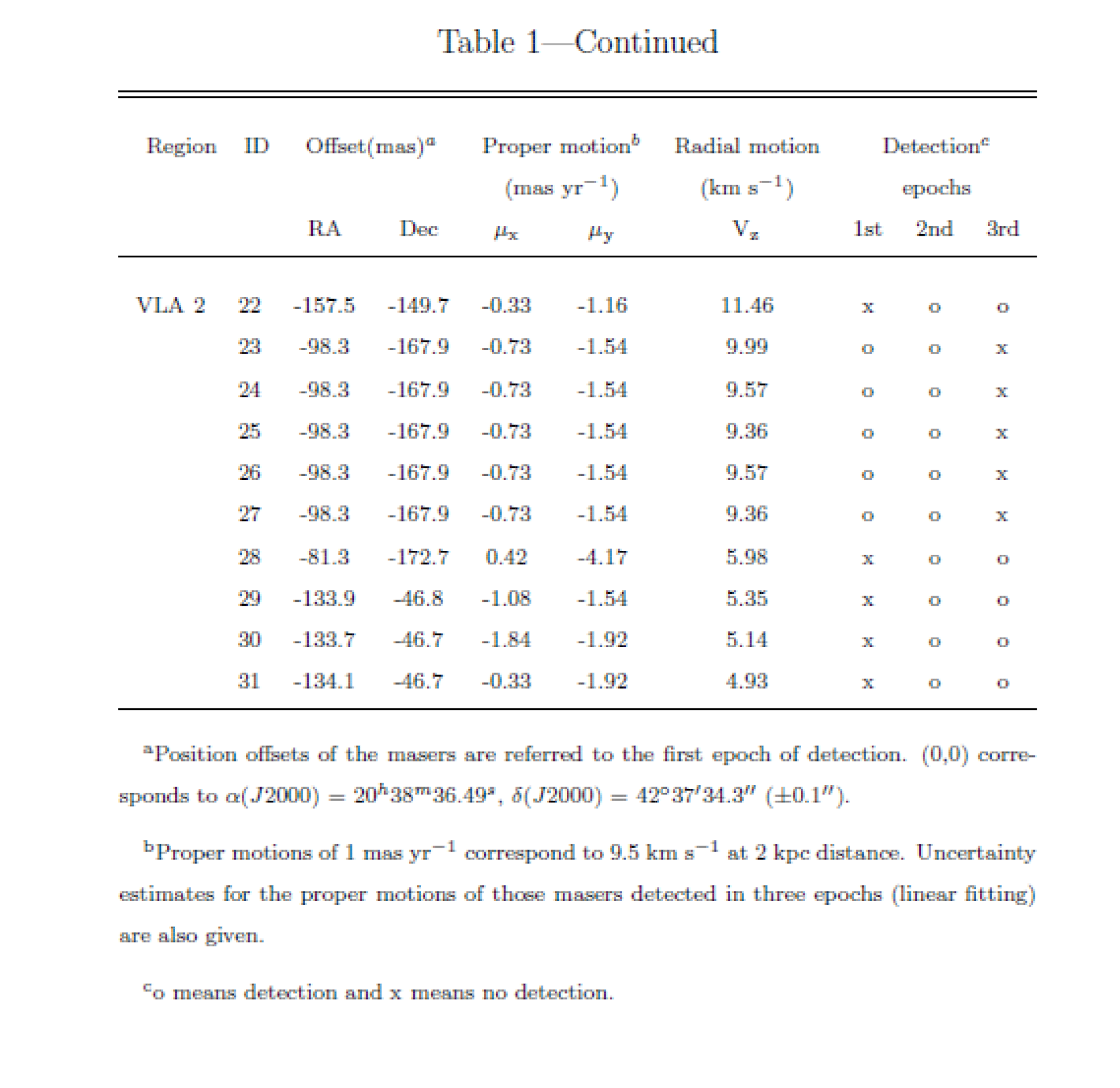}
\bigskip
\end{figure}
\clearpage

\begin{figure}
\includegraphics[scale=0.7]{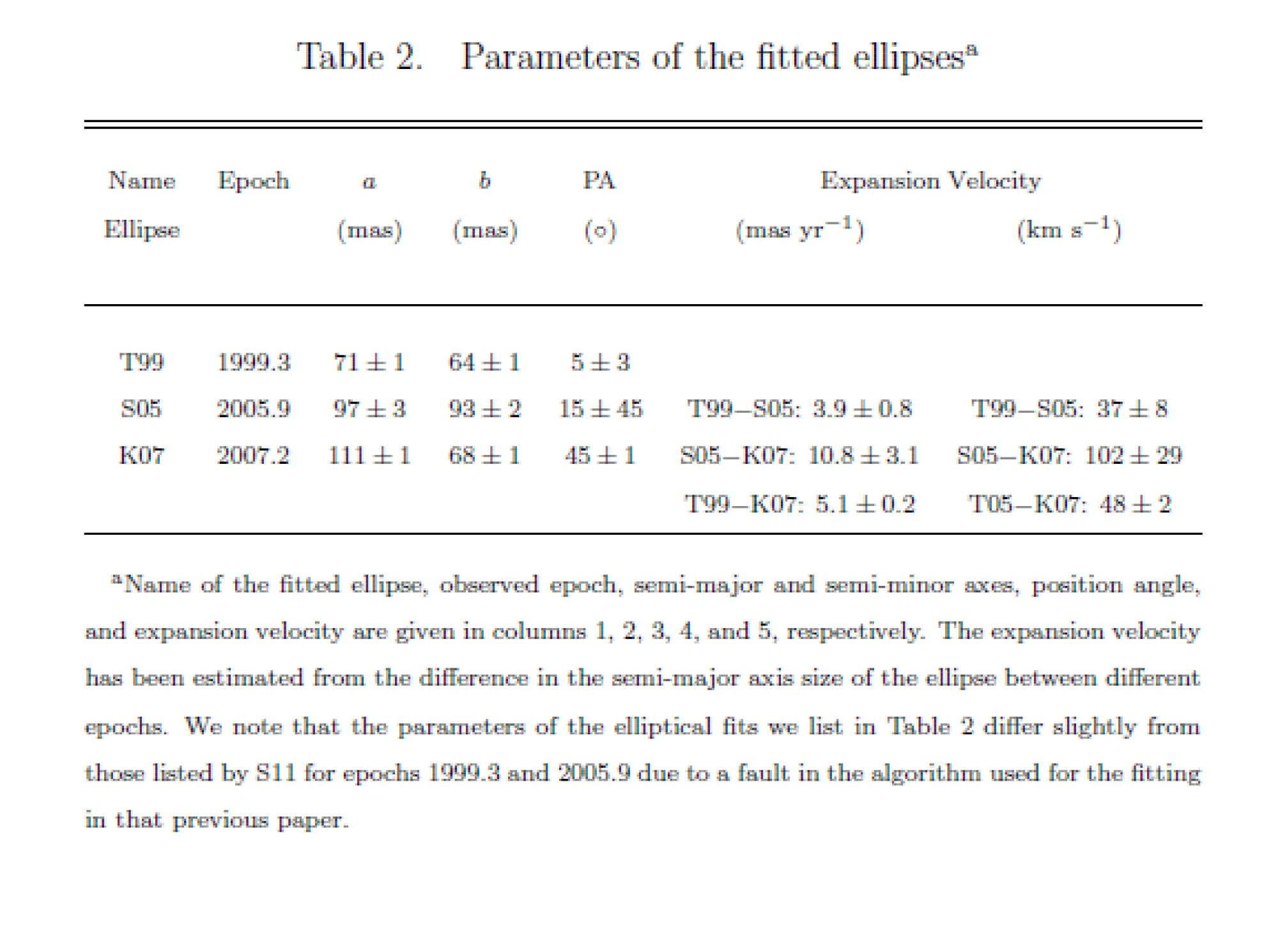}
\bigskip
\end{figure}
\clearpage

\section*{SUPPORTING MATERIAL}

Additional Supporting Information may be found in the online version of this article:

{\bf Animation.} VideoT03.wmv: Motions of the H$_2$O masers in the VLA~2 shell obtained from the VLBA data of Torrelles et al. (2003), extrapolated to ten years starting on 1999 April 2. This animation shows larger velocities of the masers along the northeast-southwest direction.

\end{document}